\documentclass[12pt]{article}
\usepackage[body={17.5cm, 22cm},right=2cm]{geometry}
\usepackage{amssymb,graphicx}
\usepackage{amsmath}
\usepackage{hyperref}

\def\e{{\rm e}}

\newcommand{\be}{\begin{equation}}
\newcommand{\ee}{\end{equation}}
\newcommand{\bea}{\begin{eqnarray}}
\newcommand{\eea}{\end{eqnarray}}
\newcommand{\bg}{\begin{gather}}
\newcommand{\eg}{\end{gather}}
\newcommand{\bseq}{\begin{subequations}}
\newcommand{\eseq}{\end{subequations}}

\begin{document}
\thispagestyle{empty}
\begin{flushright}
TCC-23-09\\
UTTG-06-09
\end{flushright}
\vskip .8cm
\begin{center}
{\LARGE\bf Signatures of a Graviton Mass \\[.3cm]
in the Cosmic Microwave Background }\\
\vspace{1.2cm}
{ \bf
Sergei~Dubovsky$^{a,b}$, Raphael~Flauger$^c$, Alexei~Starobinsky$^d$,
Igor~Tkachev$^{b}$}\\
\vspace{1.1cm}
{\small  \textit{  $^{\rm a}$ Department of Physics, Stanford University,
Stanford, CA 94305, USA }}\\
\vspace{.3cm}
{\small  \textit{  $^{\rm b}$
Institute for Nuclear Research of the Russian Academy of Sciences, \\
        60th October Anniversary Prospect, 7a, 117312 Moscow, Russia}}\\
\vspace{.3cm}
{\small  \textit{  $^{\rm c}$
Theory Group, Department of Physics, and Texas Cosmology Center\\
University of Texas at Austin, Austin, TX 78712, USA
}}\\
\vspace{.3cm}
{\small  \textit{  $^{\rm d}$
Landau Institute for Theoretical Physics, Russian Academy of Sciences,\\
119334 Moscow, Russia
}}
\end{center}
\vskip .8cm
\begin{abstract}
\vskip .5cm\noindent There exist consistent low energy effective
field theories describing gravity in the Higgs phase that allow
the coexistence of massive gravitons and the conventional $1/r$
potential of gravity. In an effort to constrain the value of the
graviton mass in these theories, we study the tensor contribution
to the CMB temperature anisotropy and polarization spectra in the
presence of a non-vanishing graviton mass. We find that the
observation of a B-mode signal consistent with the spectrum
predicted by inflationary models would provide the strongest limit
yet on the mass of an elementary particle -- a graviton -- at a
level of $m\lesssim 10^{-30}$~eV$\approx(10\mbox{ Mpc})^{-1}$. We
also find that a graviton mass in the range between $(10\mbox{
Mpc})^{-1}$ and $(10\mbox{ kpc})^{-1}$ leads to interesting
modifications of the polarization spectrum. The characteristic
signature of a graviton mass in this range would be a plateau in
the $B$-mode spectrum up to angular multipoles of $\ell\sim 100$.
For even larger values of the graviton mass the tensor
contribution to the CMB spectra becomes strongly suppressed.
\end{abstract}
\newpage
\setcounter{page}{1}
\section{Introduction}

The possibility of a nonzero graviton mass is intriguing and has
attracted the attention of theorists for a long time, see
e.g.~\cite{Fierz:1939ix,vanDam:1970vg,Zakharov,Vainshtein:1972sx,Boulware:1973my,Arkani-Hamed:2002sp}
(see \cite{Rubakov:2008nh} for a recent review). Massive gravitons
have a number of peculiar properties, such as the van
Dam-Veltman-Zakharov discontinuity~\cite{vanDam:1970vg,Zakharov}, ghost
instabilities~\cite{Boulware:1973my} and strong coupling effects
at unacceptably low energy scales~\cite{Arkani-Hamed:2002sp} which
complicate the construction of a sensible theory. However, in
light of the cosmological constant problem and coincidence
problems between baryonic matter, dark matter, and dark energy,
there has been increased interest in massive gravity theories, and
it was found that models of modified gravity with
Lorentz-violating graviton mass terms may avoid all of these
problems~\cite{Arkani-Hamed:2003uy,Rubakov:2004eb,Dubovsky:2004sg}.
The effective field theories remain valid up to reasonably large
energy scales provided a large enough subgroup of the full
diffeomorphism group is left unbroken by the graviton mass.

From a phenomenological point of view, the class of models
characterized by the residual local symmetry $x^i\to x^i+\xi^i(t)$
is of particular interest.  Among the possible choices of residual
subgroups required by consistency, this is the only choice giving
rise to massive gravitational waves~\cite{Dubovsky:2004sg}.\footnote{Note that in the Lorentz-violating theories a presence
of graviton mass terms does not yet imply that the tensor modes
are massive.  For instance, the graviton mass term
$m_{00}^2h_{00}^2$ just fixes the gauge $h_{00}=0$.} Some
phenomenological and cosmological consequences of these models
were studied in
Refs.~\cite{Dubovsky:2004ud,Dubovsky:2005dw,Bebronne:2007qh,Dubovsky:2007zi}.\footnote{A related class of bigravity models was considered in~\cite{Berezhiani:2007zf,Berezhiani:2008nr}.}
In particular, late time cosmological attractors have been found
where an additional dilatation symmetry $t\to\lambda t,\;
x^i\to\lambda^{-\gamma} x^i$ gets restored, where $\gamma$ is a
real constant. The main properties of the system in the vicinity
of these attractors can be summarized as follows.
\begin{itemize}
\item[({\it i})] The evolution of the background cosmology is
described by the usual Friedmann equation. For suitable choices of
the Lagrangian in the symmetry breaking sector that leads to the
graviton mass there will be an additional ``dark energy''
component that may contribute to the observed accelerated
expansion of the universe. \item[({\it ii})] The equations
describing the evolution of the scalar and vector fluctuations
coincide with those of General Relativity. \item[({\it iii})] The
equation describing the evolution of the tensor fluctuations is
modified by the presence of a mass term. In other words, the
dispersion relation for gravitational waves takes the form
$\omega^2=p^2+m_g^2$.
\end{itemize}

As a consequence of $(i)$ and $(ii)$, the strongest bounds on the
graviton mass, $m_g$, in this class of models come from direct or
indirect observations of gravitational waves. Since these
observations are very limited so far, this opens up the
possibility for $m_g$ to be quite large. The best constraint on
the graviton mass in these models currently comes from indirect
evidence for the emission of gravitational waves from binary
pulsars timing \cite{pulsars} yielding an upper limit of

\begin{equation}
\label{bound}
m_g\lesssim  10^4 {\rm ~pc}^{-1} \approx 3\times 10^{-15}~{\text{cm}}^{-1}
\approx 6\times 10^{-20}\,eV.
\end{equation}

Relatively large graviton masses $m_g\gtrsim 0.1 \mbox{~pc}^{-1}$
could be detected by observing the characteristic signature of a
large graviton mass, a strong monochromatic signal in
gravitational wave detectors due to relic gravitons at a frequency
equal to the graviton mass \cite{Dubovsky:2004ud}. This signal
might be observed either by  LISA  or using millisecond pulsar
timing data. (See, {\it e.g.}~\cite{Pshirkov:2008nr} for a recent analysis.)

Graviton masses close to the bound (\ref{bound}) can also be found
by using higher frequency gravitational wave detectors to measure
a time delay between optical and gravitational wave signals from a
distant source.

Finally, these theories may give rise to non-universality of high
multipoles of the galactic black hole metric. If present, these
may be detected by LISA \cite{Dubovsky:2007zi}. 

The characteristic energy scale $\Lambda$ of the symmetry breaking
sector that leads to a massive graviton is of order $\Lambda \sim
\sqrt{M_{Pl} m_g}$. This scale is of the same order as the energy
density scale of the Universe now if $m_g\sim H_0$ where
$H_0\approx 0.7 \times (3\mbox{ Gpc})^{-1}$ is the current Hubble
constant.\footnote{We will be using units such that $c=\hbar=1$
throughout.} In other words, if the observed acceleration of the
Universe were due to some modification of gravity that gives rise
to a graviton mass, one would expect the mass to be of order
$H_0$, far below the reach of the experiments mentioned above.
This motivates us to look for signatures that are sensitive to
much smaller values of the graviton mass.

One expects that a small graviton mass will leave an imprint in
the temperature anisotropy and polarization spectra of the cosmic
microwave background (CMB). We study the contribution of tensor
perturbations to the CMB spectra in a modified gravity theory with
the properties outlined above and show that this is indeed the
case. We treat the graviton mass as a phenomenological parameter.
Our results will therefore be valid for any theory in which a
modification of gravity amounts to massive gravitational waves
with the $\Lambda$CDM cosmological background being unchanged. 

The paper is organized as follows. After a summary of the basic
equations, we start with an analytic discussion of the CMB
spectrum in a massive gravity theory in subsection 2.1. After an
analytic discussion of the general properties of the B-mode
spectrum, we analytically calculate the contribution to the CMB
B-type polarization for low multipole coefficients ignoring the
effects of reionization and show that the characteristic feature
in the low $\ell$ range is a plateau. For a range of masses above
the Hubble rate at the time of recombination this contribution
even dominates over the contribution from reionization and is a
good approximation to the full numerical spectrum for low
multipole coefficients. We then calculate the contributions to the
CMB temperature anisotropy and polarization from a spatially
homogeneous tensor mode. The existence of such a contribution is a
very unusual property of massive gravity theories. In the massless
case, tensor modes are frozen as long as their wavelength is
larger than the Hubble scale $H^{-1}$ where $H(t)\equiv
a^{-1}(da/dt)$ and $a(t)$ is the scale factor of an isotropic
Friedmann-Robertson-Walker (FRW) cosmological model. Since
temperature anisotropies get generated only by a time varying
tensor mode, this implies that very long wavelengths cannot
contribute.\footnote{The solution of the zero mode equation that
does not decay at late times is in fact even a pure gauge mode in
the massless case.} If the graviton is massive, however, when the
expansion rate of the universe drops below the graviton mass,
these modes acquire an oscillatory time dependence with a
frequency set by the graviton mass, and a decreasing amplitude due
to Hubble friction. As a consequence, long wavelength modes in a
massive gravity theory will generate a temperature anisotropy
quadrupole that will get converted into a polarization quadrupole
during recombination if the mass is in the right range and more
efficiently once the universe becomes reionized. The zero mode
does not contribute to the $B$-mode spectrum, but does contribute
to the temperature $T$, polarization $E$-mode and $TE$
cross-correlation quadrupoles.

We then proceed to a numerical treatment in subsection 2.2. The
most interesting result is that a graviton mass would strongly
modify the shape of the $B$-mode spectrum for $\ell<100$. If such
modifications are observed in experiments such as CMBPol, it would
provide strong support for massive gravity theories. If on the
other hand the observed spectrum is consistent with General
Relativity, this would imply an upper bound on the graviton mass
of $m_g<(10 Mpc)^{-1}\approx 10^{-30} eV$. We conclude with a
brief summary of the signatures of a graviton mass in the CMB in
Section 3.

\section{Tensor Contribution to the CMB for a Massive Graviton}
Since its discovery about 45 years ago, the cosmic microwave
background radiation has greatly improved our understanding of the
very early universe. With the help of current and future
experiments this trend is likely to continue. Especially the
detection of a B-mode signal would provide valuable information
that could be used to verify and explore the details of inflation.
The B-mode signal would also provide us with a powerful tool to
detect or constrain cosmic strings, and, as we shall see, would
also help us put tight bounds on massive gravity theories. In the
near future the best constraints on or possible detections of a
B-mode signal are expected to come from experiments such as
Spider, PolarBeaR, EBEX, BICEP/SPUD, $C_\ell$OVER, from the Planck
satellite, and in the slightly more distant future hopefully from
a mission like CMBPol~\cite{Baumann:2008aq}.

The properties of the early universe are encoded in correlations
of the temperature anisotropies and polarization patterns at
different points in the sky. The quantities most commonly used to
represent the two-point correlations are the TT as well as the TE,
EE, and BB multipole coefficients. The contribution of the tensor
fluctuations to them is given by

\begin{eqnarray}
&&C^T_{BB,\ell}= \pi^2T_0^2\int_0^\infty q^2\,dq \nonumber\\ &&~~~~
\times\left|\int_{\tau_1}^{\tau_0}d\tau\,P(\tau)\,\Psi(q,\tau)\,\left\{\left[ 8\rho+ 2\rho^2\frac{\partial}{\partial\rho}\right]
\frac{j_\ell(\rho)}{\rho^2}\right\}_{\rho=q(\tau_0-\tau))}\right|^2\;,\label{eq:bbex}\\
&&C^T_{EE,\ell}= \pi^2T_0^2\int_0^\infty q^2\,dq \nonumber\\ &&~~~~
\times\left|\int_{\tau_1}^{\tau_0}d\tau\,P(\tau)\,\Psi(q,\tau)\,\left\{\left[12+8\rho\frac{\partial}{\partial\rho}-\rho^2+\rho^2\frac{\partial^2}{\partial\rho^2}\right]
\frac{j_\ell(\rho)}{\rho^2}\right\}_{\rho=q(\tau_0-\tau))}\right|^2\;,\label{eq:eeex}\\
&&C^T_{TE,\ell}=-2\pi^2 T_0^2\sqrt{\frac{(\ell+2)!}{(\ell-2)!}}\int_0^\infty q^2\,dq\nonumber\\&&~~~~\times\int_{\tau_1}^{\tau_0}d\tau\,P(\tau)\Psi(q,\tau)\left\{\left[12+8\rho\frac{\partial}{\partial\rho}-\rho^2+\rho^2\frac{\partial^2}{\partial\rho^2}\right]
\frac{j_\ell(\rho)}{\rho^2}\right\}_{\rho=q(\tau_0-\tau)}\nonumber\\&&~~~~\times
\int_{\tau_1}^{\tau_0}d\tau'\,d(q,\tau')
\left\{ \frac{j_\ell\Big(q(\tau_0-\tau')\Big)}{q^2(\tau_0-\tau')^2}\right\}\;,\label{eq:teex}\\
&&
C^T_{TT,\ell}=\frac{4\pi^2(\ell+2)!T_0^2}{(\ell-2)!}\int_0^\infty q^2\,dq\left|\int_{\tau_1}^{\tau_0} d\tau\;d(q,\tau)\,\frac{j_\ell\Big(q(\tau_0-\tau)\Big)}{q^2 (\tau_0-\tau)^2}\right|^2\;.\label{eq:ttex}
\end{eqnarray}

These formulas are equivalent to those of Zaldarriaga and Seljak
\cite{seljak} up to integration by parts.\footnote{We also use
slightly different conventions. Their gravitational wave amplitude
$h$ and power spectral function $P_h(k)$ are related to our
gravitational wave amplitude ${\cal D}_q(\tau)$ by
$h\sqrt{P_h}={\cal D}/2$.  In consequence, their function
$\Psi\sqrt{P_h}$ is $1/4$ times our source function $\Psi$.} While
this makes no difference in the massless case, in the massive case
the formulas as written here are better suited for numerical
calculations.

In these equations, $q=pa(t)$ is the comoving momentum; $\tau=\int
\, dt/a(t)$ is the conformal time; $T_0=2.725\,K$ is the microwave
background temperature at the present conformal time $\tau_0$;
$P(\tau)=\dot\kappa\exp[-\int_\tau^{\tau_0}\dot\kappa(\tau')\,d\tau']$
is the  probability distribution of last scattering (or visibility
function), with $\dot\kappa(\tau)$ the photon collision frequency
(or differential optical depth); $\tau_1$ is any time taken early
enough before recombination so that any photon present at $\tau_1$
would have collided many times before the present; and
$\Psi(q,\tau)$ is the ``source function,'' which is customarily
calculated from a hierarchy of equations for partial-wave
amplitudes~\cite{polnarev,Crittenden:1993ni} (the dot means the
derivative with respect to $\tau$):
\begin{eqnarray}
&&\dot{\tilde\Delta}^{(T)}_{T,\ell}(q,\tau)+\frac{q}{(2\ell+1)}\Big((\ell+1)\tilde\Delta^{(T)}_{T,\ell+1}(q,\tau)-\ell \tilde\Delta^{(T)}_{T,\ell-1}(q,\tau)\Big)
\nonumber\\&&~~~~=\Big(-2\dot{\cal D}_q(\tau)+\dot\kappa(\tau)\Psi(q,\tau)\Big)\,\delta_{\ell,0}-\dot\kappa(\tau)\tilde\Delta^{(T)}_{T,\ell}(q,\tau)\;,\label{eq:bt}\\[0.3cm]&&
\dot{\tilde\Delta}^{(T)}_{P,\ell}(q,\tau)+\frac{q}{(2\ell+1)}\Big((\ell+1)\tilde\Delta^{(T)}_{P,\ell+1}(q,\tau)-\ell \tilde\Delta^{(T)}_{P,\ell-1}(q,\tau)\Big)
\nonumber\\&&~~~~=-\dot\kappa(\tau)\Psi(q,\tau)\,\delta_{\ell,0}-\dot\kappa(\tau)\tilde\Delta^{(T)}_{P,\ell}(q,\tau)\;,\label{eq:bp}
\end{eqnarray}
with
\begin{eqnarray}
&&\Psi(q,\tau)=\frac{1}{10}\tilde\Delta^{(T)}_{T,0}(q,\tau)+\frac{1}{7}\tilde\Delta^{(T)}_{T,2}(q,\tau)+
\frac{3}{70}\tilde\Delta^{(T)}_{T,4}(q,\tau)-\frac{3}{5}\tilde\Delta^{(T)}_{P,0}(q,\tau)\nonumber\\&&~~~~~~~+\frac{6}{7}\tilde\Delta^{(T)}_{P,2}(q,\tau)-\frac{3}{70}\tilde\Delta^{(T)}_{P,4}(q,\tau)\;.
\end{eqnarray}
Alternatively, the source function can be calculated from an integral equation~\cite{Weinberg:2006hh,Baskaran:2006qs}, and we have used both approaches.

Here ${\cal D}_q(\tau)$ is the gravitational wave amplitude (apart from terms that decay outside the horizon), defined by
\begin{equation}
\delta g_{ij}({\bf x},\tau)=a^2\sum_\pm\int d^3q\;e^{i{\bf q}\cdot{\bf x}}\,\beta({\bf q},\pm 2)\,e_{ij}(\hat{q},\pm 2)\,{\cal D}_q(\tau)\;,
\end{equation}
with $\beta({\bf q},\pm 2)$ and $e_{ij}(\hat{q},\pm 2)$ the stochastic parameter and polarization tensor for helicity $\pm 2$, normalized so that
\begin{equation}
\langle \beta({\bf q},\lambda)\,\beta^*({\bf q}',\lambda')\rangle=\delta_{\lambda\lambda'}\delta^3({\bf q}-{\bf q}')\;,
\end{equation}
and for $\hat{q}$ in the 3-direction
\begin{equation}
e_{11}(\hat{q},\pm 2)=-e_{22}(\hat{q},\pm 2)=1/\sqrt{2}\;,~~~
e_{12}(\hat{q},\pm 2)=e_{21}(\hat{q},\pm 2)=\pm i/\sqrt{2}\;.
\end{equation}
Finally, $d(q,\tau)$ is the quantity
\begin{equation}
d(q,\tau)\equiv \exp\left[-\int_\tau^{\tau_0}d\tau'\,\dot\kappa(\tau')\right]\,\left(\dot{\cal D}_q(\tau)-\frac{1}{2}\dot\kappa(\tau)\Psi(q,\tau)\right)\;.
\end{equation}


In massive gravity, the evolution of the gravitational wave
amplitude is described by the solution to the equation for a
minimally coupled {\it massive} scalar
field~\cite{Dubovsky:2004ud,Dubovsky:2005dw}\footnote{In writing
this equation, we drop contributions on the right hand side due to
anisotropic stress generated by neutrinos and
photons~\cite{AZakharov,Weinberg:2003ur}. This is done merely for
simplicity and we will include these effects in our calculations.
} \be \label{heq} \ddot{\cal D}_q(\tau)+2{\frac{\dot
a}{a}}\dot{\cal D}_q(\tau)+(q^2+m_g^2a^2){\cal D}_q(\tau)=0 \; .
\ee

In conventional cosmological perturbation theory, the solution of
Eq.~(\ref{heq}) which remains finite for $a\to 0$ and has a
wavelength larger than the Hubble scale $H^{-1}$  is not
observable locally and does not contribute to the CMB anisotropy
and polarization spectra. On the other hand, the other, decaying
graviton mode produces locally measurable effects even in the
$k\to 0$ limit. Since it does decay, it is usually assumed to be
negligible by the time of recombination.

In the massive gravity case, Eq.~(\ref{heq}) implies that a
homogeneous metric perturbation starts to oscillate with a
frequency equal to $m_g$ when the expansion rate $H$ drops below
the graviton mass $m_g$. This is completely analogous to what
happens for light scalar fields ({\it e.g.} axion, moduli, \dots).
An important difference, however, is that in the case of the
graviton these oscillations may directly affect the CMB spectra
(or, if the mass $m_g$ is high enough, may be observed by
gravitational wave detectors), similar to a super-Hubble decaying
mode or a generic sub-Hubble tensor perturbation in the massless
limit. Indeed, the presence of a zero mode implies that,
superimposed upon a conventional Hubble expansion, spatial metric
components experience anisotropic (but homogeneous) high frequency
oscillations with a small amplitude. The effect of such
oscillations on the CMB spectra can easily be understood
analytically and we will return to this at the end of subsection
2.1.

\subsection{Analytic results for low multipole coefficients}
\label{zero_mode} Before presenting the results of the numerical
calculations in the next subsection, let us start with a brief
analytic discussion of the spectrum. For the most part, we will
limit ourselves to the contribution to the spectrum generated
during recombination and ignore the effects of reionization. In
the massless case, the effects of reionization give the dominant
contribution to the spectrum for $\ell\lesssim 20$ but leave the
higher multipole coefficients unchanged (or rather change them
trivially by an overall rescaling by $e^{-2\tau_\text{reion}}$,
where $\tau_\text{reion}=0.087\pm0.017$ is the optical depth of
the medium due to reionization and despite the clash of notation
should not be confused with conformal time), so that ignoring
the effects of reionization is good as long as one is interested
in $\ell>20$. In the massive case, our numerical results indicate
that for a range of masses above the Hubble rate at recombination
the contribution from recombination provides a good approximation
to the spectrum even at low $\ell$ providing additional motivation
for this simplifying assumption.

We will focus on the B-mode spectrum because it is the most
interesting one from an experimental point of view. The discussion
could be straightforwardly  extended to include the TT, TE, and EE
spectra, but we  limit ourselves to the contribution of the zero
mode to those.

Depending on their comoving momentum, the modes fall into one of
two classes or one of three classes depending on whether the mass
is smaller than the Hubble rate at recombination or larger than
that.

For masses below the Hubble rate at recombination, the first
possibility is that modes are relativistic at the time they enter
the horizon. In this case they will still be relativistic during
recombination. These modes are essentially unaffected by the
graviton mass, and the spectrum for the values of $\ell$ these
modes contribute to is expected to agree with the one in the
massless case. The second possibility is that the modes enter the
horizon when they are already non-relativistic. The multipole
coefficients these modes contribute to will be different from the
ones for the massless case and we will discuss those in more
detail below.

For masses above the Hubble rate at recombination there is a third
option. The modes can be relativistic as they enter the horizon
but become non-relativistic by the time of recombination.

\begin{figure}[t]
\begin{center}
\includegraphics[width=5.3in]{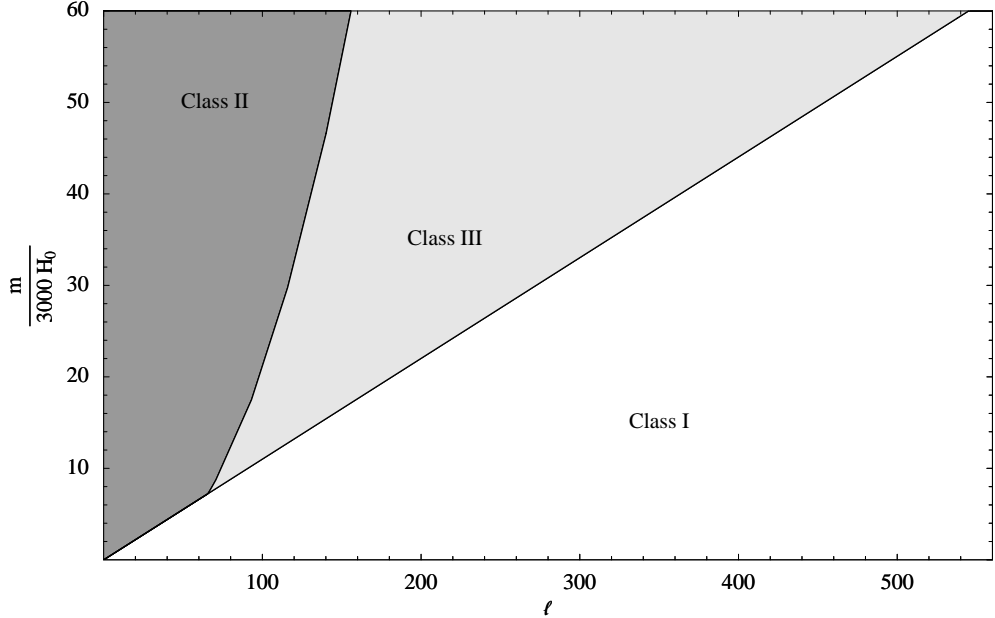}
\end{center}
\caption{This plot summarizes the different behaviors of modes and
which range of multipole coefficients they contribute to. Class I
corresponds to modes that are relativistic at recombination. Class
II corresponds to modes that are non-relativistic as they enter
the horizon and during their subsequent evolution. Class III
corresponds to modes that enter the horizon when they are
relativistic but become non-relativistic before recombination.}
\label{fig:classes}
\end{figure}

The results are summarized in terms of the range of multipole
coefficients the different classes affect for a given mass in
Figure~\ref{fig:classes}.

Let us now discuss these regimes in more detail. Both for masses
below and above the expansion rate during recombination, the
short-wavelength modes that are in the relativistic regime during
recombination, \be \label{tomasslessk}
{\frac{q}{a(\tau_r)}}>m_g\;, \ee are not affected by the graviton
mass and will lead to the same spectrum as in the massless case.
In terms of the multipole number $\ell\approx q
(\tau_0-\tau_{rec})$ this transition to the massless regime
corresponds to \be \label{tomassless} \ell\gg \ell_0\equiv m_g
a(\tau_r) (\tau_0-\tau_r)\approx {\frac{m_g}{H_0}
(1+z_r)^{-1}}\int_{(1+z_r)^{-1}}^1{\frac{dx}{\sqrt{\Omega_\Lambda
x^4+\Omega_m x+\Omega_r}}}\approx 3.3(1+z_r)^{-1}
{\frac{m_g}{H_0}}\;, \ee where $z_r\approx 1088$ is the redshift
at recombination, and we have used the five-year WMAP values for
the cosmological parameters~\cite{Komatsu:2008hk}. In particular,
equation (\ref{tomassless}) implies that, ignoring the contribution
generated during reionization, the $B$-mode spectrum is not
modified for masses smaller than $\sim 300 H_0$, which is the
scale corresponding to the size of the visible patch of the
Universe during recombination.

 For larger masses, but still smaller than the expansion rate at
recombination,\linebreak \mbox{$m_g<H(\tau_r)\approx 2\times 10^4 H_0$}, the modes
that are affected by the graviton mass are superhorizon during
recombination because they satisfy $\frac{q}{a(\tau_r)}\lesssim
m_g<H(\tau_r)$. As a consequence these modes do not oscillate
during recombination. Nevertheless, just like in the massless
case, the corresponding source term $\dot{\cal D}_q$ in
(\ref{eq:bt}) is non-zero (though small) and some amount of
polarization is being generated. In contrast to the massless case,
however, where the value of $\dot{\cal D}_q$ at recombination is
determined by the value of $q$ alone, and goes to zero as $q\to
0$, in the massive case $\dot{\cal D}_q$   depends on
$(q^2+m_g^2a(\tau_r)^2)$ and is independent of $q$ for long
wavelength modes leading to an enhancement of the spectrum for
$\ell<\ell_0$.

For values of the graviton mass larger than the Hubble rate at
recombination, $m_g\gtrsim H(\tau_r)$, all modes start to
oscillate before recombination. The modes with long wavelengths
start to oscillate as soon as the expansion rate of the universe
drops below the graviton mass, {\it i.e.} at a time
$\tau_m<\tau_r$, such that \be \label{longosc} H(\tau_m)=m_g\;.
\ee In particular, \be \label{sol} {\cal D}_q\simeq {\cal
D}_{q0}\frac {\sin m_gt}{m_gt}\ee for $(q/a)^{2}\ll m_gH$ at the
matter dominated stage, where $t\propto \tau^3$. Shorter modes
start to oscillate when they enter the horizon just like in the
massless case. The transition between these two regimes happens at
$q_m=m_g a(\tau_m)$.

To a good approximation, all modes with momenta smaller than $q_m$
have the same evolution--they are frozen until $\tau_m$, and
oscillate afterwards with a frequency set by the mass. This value
of comoving momentum, $q_m$, corresponding to the transition
between class II and III translates to a value in $\ell$-space of
\be \ell_m=3.3(1+z_m)^{-1}{\frac{m_g}{ H_0}}\;. \ee Here $z_m$ is
the redshift corresponding to $\tau_m$ and is determined by
condition (\ref{longosc}) which can be written more explicitly as
\be \label{zm} H_0\sqrt{\Omega_m (1+z_m)^3+\Omega_r
(1+z_m)^4}=m_g\;. \ee At $m_g=H(\tau_r)$ the multipole number
$\ell_m$  coincides with $\ell_0$ and takes a value of around
$\ell_m\sim 65$. At higher masses these two scale are different.
According to (\ref{tomassless}) $\ell_0$ grows linearly with mass.
On the other hand, $\ell_m$ grows more slowly, $\ell_m\propto
m_g^{1/3}$ for masses that become relevant during matter
domination ({\it i.e.}, for $m_g\lesssim 1.5\times 10^5 H_0$), and
$\ell_m\propto m_g^{1/2}$ at higher $m_g$. For masses much larger
than this, all modes oscillate rapidly during recombination. As a
consequence the polarization signal gets averaged out and becomes
strongly suppressed.

Modes corresponding to angular scales between $\ell_m$ and
$\ell_0$ enter the horizon when they are still relativistic, but
become non-relativistic before recombination. As a result, they
are still expected to exhibit the conventional oscillation pattern
in the angular spectrum, but the phase of oscillations is
different because the oscillations at late times are driven by the
mass rather than the spatial momentum.

After this discussion of various regimes, let us take a more
detailed look  at the spectrum. As discussed, for the modes
referred to as class I in Figure~\ref{fig:classes} that are
relativistic during recombination the spectrum to a good
approximation agrees with the one in the massless case. While a
number of analytic results for the temperature anisotropy and
polarization have been found for this case, we will not review
those here and refer the interested reader to the
literature~\cite{St85,Zaldarriaga:1995gi,Pritchard:2004qp,Keating:2006zy,Flauger:2007es}.

As can be seen by inspection of equation~\eqref{heq}, for the
modes referred to as class II in Figure~\ref{fig:classes} the
dependence of the gravitational wave amplitude on comoving
momentum is trivially given by that of the power spectrum because
they are frozen as long as they are outside the horizon and the
mass already dominates by the time they enter. This does not in
general guarantee that the same is true for the source function as
it will generically develop its own $q$-dependence. To a good
approximation the momentum dependence of the source function
during recombination is the same as that of the gravitational wave
amplitude provided the comoving momentum of the mode is less than
the duration of recombination in conformal time, {\it i.e.}
$q\Delta \tau_\text{rec}\ll 1$. This is satisfied for all modes in
class II for the range of masses we are interested in and hence
does not provide an additional constraint.

\begin{figure}[t]
\begin{center}
\includegraphics[width=5.3in]{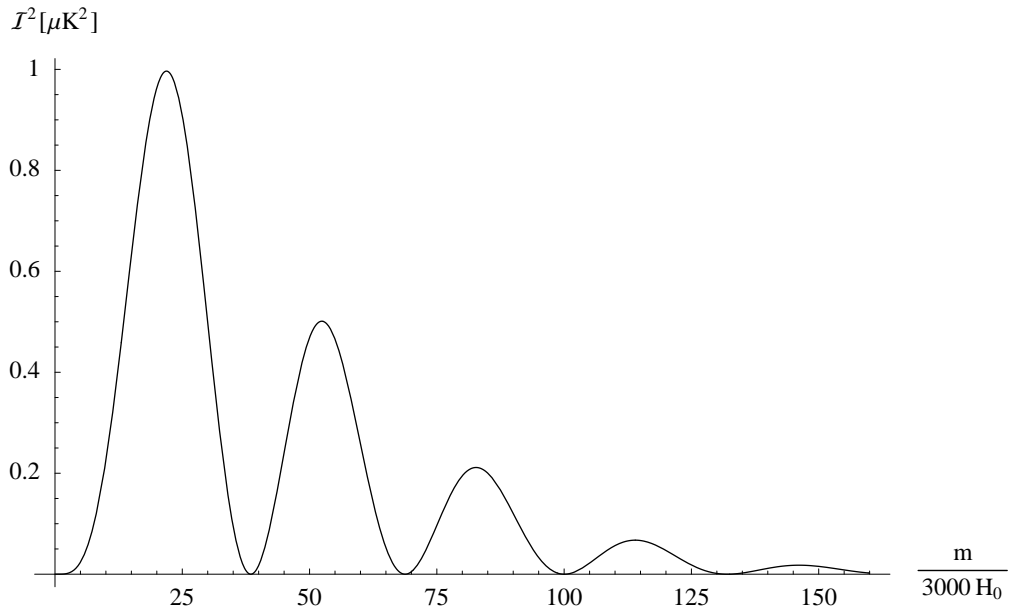}
\end{center}
\caption{This plot shows the quantity ${\cal I}^2$ in $(\mu K)^2$
as a function of mass for a scalar amplitude of $\Delta_{\mathcal
R}^2=2.41\times 10^{-9}$, a tensor-to-scalar ratio of $r=1$, and a
tensor spectral index $n_T=0$.} \label{fig:I2}
\end{figure}

For modes in class II, the dependence of the gravitational wave
amplitude and that of the source function on comoving momentum are
then trivially given by that of the power spectrum, implying {\it
e.g.} for a standard inflationary scenario that
$\Psi(q,\tau)q^{\frac32-\frac{n_T}{2}}$ is $q$-independent. This
allows us to evaluate the expression for $C^T_{BB,\ell}$ given by
equation~\eqref{eq:bbex} analytically. Conventionally, one first
evaluates the integrals over conformal time and then integrates
over momentum. For us it will be more convenient to perform the
integral over momentum {\it first}. This is possible by rewriting
the square of the integral over time as an integral in a plane and
using the identity:
\begin{equation}
\left(8\rho+2\rho^2\frac{\partial}{\partial \rho}\right)\frac{j_\ell(\rho)}{\rho^2}=\frac{\sqrt{2\pi}}{\rho^{\frac32}}\left((2+\ell)J_{\ell+\frac12}(\rho)-\rho J_{\ell+\frac32}(\rho)\right)\,,
\end{equation}
where $J_\nu(\rho)$ is the Bessel function of the first kind. The
resulting four integrals over $q$ can then be done exactly using
an integral known as the Weber-Schafheitlin integral (see {\it
e.g.}~\cite{bateman}). Dropping terms of order
\begin{equation}\label{eq:2ndmoment}
\left\langle\frac{(\tau-\tau_L)^2}{{\tau_L}^2}\right\rangle\equiv\int\limits_{\tau_1}^{\tau_0}d\tau\, P(\tau)\Psi(q,\tau)q^{\frac32-\frac{n_T}{2}} \frac{(\tau-\tau_L)^2}{{\tau_L}^2}\,,
\end{equation}
and higher in the terms in the integrals over conformal time arising from the integral over comoving momentum, and
setting $n_T=0$ for simplicity, one finds the following expression for the power spectrum:
\begin{equation}
\frac{\ell(\ell+1)}{2\pi}C^T_{BB,\ell}=\frac{2(\ell(\ell+1)+16)}{3(\ell+2)(\ell-1)}{\cal I}^2\,.
\end{equation}
The $\ell$-dependence is now explicit and it is easy to see that for
$\ell\gtrsim 10$ this becomes independent of $\ell$. The $\ell$-independent
quantity $\cal I$ is defined as\footnote{Recall that $T_0=2.725\,K$ is the CMB temperature at the present time.}
\begin{equation}
{\cal I}=\sqrt{\frac{\pi}{2}}{T_0}\int\limits_{\tau_1}^{\tau_0}d\tau\, P(\tau)\Psi(q,\tau)q^{3/2}\,,
\end{equation}
and it encodes the dependence of the spectrum on the mass through the dependence of the source function on the mass. The quantity
${\cal I}^2$ as a function of mass is shown in Figure~\ref{fig:I2}
for a scalar amplitude of $\Delta_{\mathcal R}^2=2.41\times
10^{-9}$, a tensor-to-scalar ratio $r=1$.

The oscillatory features seen in the plot can be understood from
the fact that the tensor perturbations of the metric take the form
given in equation~\eqref{sol}. In particular, they are regular in the limit $t\to
0$ and the ``decaying" mode, which diverges in this limit, is absent. In turn,
this property (which also takes place for larger values of $q$,
$(q/a)^2 \ge m_gH$) is a consequence of local isotropy of the
Universe at very early times. We assume the latter to be produced
by inflation and use the inflationary prediction for the
primordial power spectrum, but the existence of the oscillations
in $\cal I$, as well as the existence of oscillations in the
multipole power spectra of polarization and temperature anisotropy
seen in Figures \ref{fig:clbb} and \ref{fig:4m10} below
("primordial peaks", similar to the well known acoustics peaks
produced by scalar perturbations but with approximately twice less
asymptotic period in $\ell$, $T_\ell=\pi(\tau_0-\tau_r)/\tau_r\approx
140$~\cite{PS96}), is a more general phenomenon not depending on
how this early time isotropy was achieved.

As we will derive shortly, what enters into the source function in
a crucial way is the tensor perturbation of the metric evaluated
at the time of recombination. This quantity viewed as a function
of the graviton mass oscillates around zero, implying that the
integral does, too. After squaring, this will give rise just to
what is seen in Figure~\ref{fig:I2}.

Ignoring terms higher order in the quantity~\eqref{eq:2ndmoment}
is typically only a good approximation for $\ell\lesssim 30$, but
our numerical calculations show that the plateau persists to
higher values of $\ell$. For the values of masses where ${\cal I}$
is close to zero, the higher order terms in this expansion give
the leading contribution and have to be included even for low
$\ell$. Furthermore, for these ranges of masses the effects of
reionization become important. Especially for masses below the
expansion rate at recombination, reionization is important as it
will increase the sensitivity of polarization measurements for
masses near the low end of the accessible mass range $m_g\sim 300
H_0$ by about one order of magnitude to $m_g\sim 30 H_0$.

After discussing the B-mode signal in a massive gravity theory,
let us briefly discuss another unusual feature in these theories,
a contribution of modes of extremely long wavelength to the TT,
EE, and TE quadrupoles. In the limit of vanishing momentum, the
Boltzmann hierarchy, {\it i.e.}
equations~\eqref{eq:bt},~\eqref{eq:bp} become very simple. Only
$\tilde\Delta^{(T)}_{T,0}(q,\tau)$ and
$\tilde\Delta^{(T)}_{P,0}(q,\tau)$ get generated while all others
remain zero. Similar equations were considered in
Ref.~\cite{baskopolnarev}, where CMB polarization and anisotropy
in the Kasner universe were studied. To calculate the contribution
to the CMB temperature anisotropy and polarization in this limit,
it is convenient to write equations~\eqref{eq:bt},~\eqref{eq:bp}
as
\begin{eqnarray}
&&\frac{d}{d\tau}\left(e^{-\kappa(\tau)}\tilde\Delta^{(T)}_{T,0}(\tau)\right)=-2d(\tau)\,,\\
&&\frac{d}{d\tau}\left(e^{-\kappa(\tau)}\tilde\Delta^{(T)}_{P,0}(\tau)\right)=-P(\tau)\Psi(\tau)\,,
\end{eqnarray}
where we have defined the integral optical depth $\kappa(\tau)$ as
\[
\kappa(\tau)=\int_\tau^{\tau_0} d\tau'\,\dot\kappa(\tau')\; .
\]

From equations~\eqref{eq:eeex},\eqref{eq:teex},
and~\eqref{eq:ttex} we see that the contribution of the zero mode
to the TT, EE, and TE quadrupoles is thus given by
\[
C^T_{TT,2}={\frac{2\pi} {75}}{T_0}^2\tilde\Delta^{(T)}_{T,0}(\tau_0)^2\;,
\]
\[
C^T_{EE,2}={\frac{4\pi}{25}}{T_0}^2\tilde\Delta^{(T)}_{P,0}(\tau_0)^2\;,
\]
and
\[
C^T_{TE,2}= -{\frac{2\pi} {25}}\sqrt{\frac23}{T_0}^2\tilde\Delta^{(T)}_{T,0}(\tau_0)\tilde\Delta^{(T)}_{P,0}(\tau_0).
\]
It is straightforward to solve equations~\eqref{eq:bt}
and~\eqref{eq:bp} for $\tilde\Delta^{(T)}_{T,0}(\tau_0)$ and
$\tilde\Delta^{(T)}_{P,0}(\tau_0)$ in this simple case. The result
is
\begin{gather}
\label{asol}
\tilde\Delta^{(T)}_{T,0}(\tau_0)=-\frac{6I_1}{7}  - \frac{I_2}{7}\;, \\
\label{bsol}
\tilde\Delta^{(T)}_{P,0}(\tau_0)=-\frac{I_1}{7}  + \frac{I_2}{7} \;,
\end{gather}
where $I_1$ and $I_2$ are the following integrals
\begin{gather}
\label{I1}
I_1=2\int_0^{\tau_0}d\tau\, \e^{-\kappa(\tau)}\dot {\cal D}(\tau)\;,\\
\label{I2}
I_2=2\int_0^{\tau_0}d\tau\, \e^{-\frac{3}{10}\kappa(\tau)}\dot {\cal D}(\tau)\;.
\end{gather}

It is convenient to discuss the behavior of the integrals
(\ref{I1}), (\ref{I2}) first neglecting the contribution from the
reionization epoch. In the absence of reionization, the functions
$\e^{-\kappa(\tau)},\;  \e^{-\frac{3}{10}\kappa(\tau)}$ have a
step-like shape and change their values from 0 to 1 at the time of
recombination, $\tau=\tau_r$. Let $a(\tau_r) \Delta \tau\sim
35$~kpc be the characteristic width of these step functions (or
the duration of the recombination epoch). The relevant parameter
which determines the behavior of integrals (\ref{I1}), (\ref{I2})
is then
\[
\delta=m_g a(\tau_r) \Delta \tau\;.
\]
For small masses, such that $\delta \ll 1$ (but, of course,
assuming $m_g>H_0$) one has \be \label{Inoreion} I_1=I_2=-2{\cal
D}(\tau_r)\; . \ee Consequently, in this case polarization is
negligible, while $\tilde\Delta^{(T)}_{T,0}(\tau_0)=2{\cal
D}(\tau_r)$ gives the temperature anisotropy quadrupole. Note that
for a fixed initial amplitude of the metric perturbation, the
anisotropy is smaller for larger $m_g$. For large masses, $\delta
\gg 1$, the mode oscillates rapidly even during recombination. As
a result, both integrals $I_1$ and $I_2$ are very small ($\propto
\e^{-\delta}$) and both temperature and polarization quadrupoles
are negligible.

The largest amount of polarization is generated for $\delta\sim
1$. In this case recombination cannot be treated as instantaneous,
so that there is no cancellation between the two terms in the
expression for the polarization
$\tilde\Delta^{(T)}_{P,0}(\tau_0)$. On the other hand, metric
oscillations during recombination do not wash out the whole effect
yet, and one gets comparable contributions to polarization and
temperature anisotropy.

Finally, let us include the effect of reionization. In the
presence of reionization Eq.~(\ref{Inoreion}) does not hold even
for small masses. Instead, one has\footnote{We assume that the
graviton mass is still large enough, $m_g\gg H_0$, so that the
graviton oscillates rapidly during the reionization. }
\begin{gather}
I_1=-2\e^{-\tau_\text{reion}}{\cal D}(\tau_r)\;,    \\
I_2=-2\e^{-\frac{3}{10}\tau_\text{reion}}{\cal D}(\tau_r)\;, 
\end{gather}
where $\tau_\text{reion}$ is again the optical depth of the medium
due to reionization. As a result, both integrals get somewhat
suppressed. On the other hand, the two terms in Eq.~(\ref{bsol})
no longer cancel, and the contributions to polarization and
temperature anisotropy can be of the same order.

\subsection{Numerical results}

To calculate the angular power spectra for values of $\ell>50$, we
use CMBfast \cite{seljak,cmbfast} with the evolution equation for
the tensor perturbations modified according to Eq.~(\ref{heq}).
For the low multipole coefficients, $\ell\leq50$, we use the
CMBfast source function, but perform the line of sight integration
in Mathematica using
equations~\eqref{eq:bbex},~\eqref{eq:eeex},~\eqref{eq:teex},
and~\eqref{eq:ttex} because the CMBfast results become unreliable
for $\ell\leq 50$ at least for large graviton masses.

The issue arises because of rapid oscillations of the source
function for all values of comoving momentum. After the
integration by parts as implemented in CMBfast, eq.~(29) of
\cite{Seljak:1996is} involves second derivatives of the source
function, which are unpleasant to deal with numerically. As a
result, the line of sight integration, as implemented in CMBfast,
produces unreliable results. The problem exists for the BB
spectrum as well but is especially severe for the low-$\ell$ parts
of the EE and TE spectra, where it appears at masses of order
$m_g\sim 3000 H_0$.

\begin{figure}[t!]
\begin{center}
\includegraphics[width=6.9in]{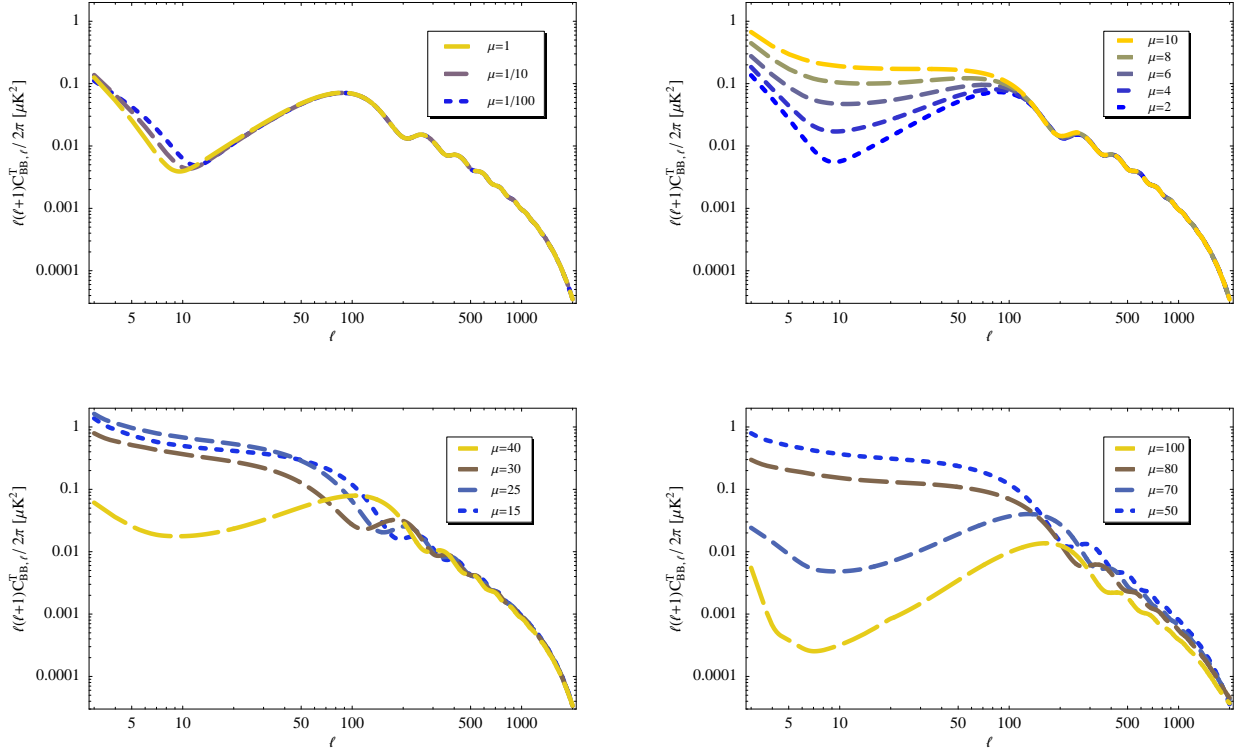}
\end{center}
\caption{These plots show the $C^T_{BB,\ell}$ multipole
coefficients for the range of masses that lead to the most
interesting signal in the CMB. The masses are given by
$m_g=\mu\times 3000 H_0$, where $\mu$ is given in the legend.
Longer dashes correspond to larger mass. All plots are for a
scalar amplitude $\Delta_{\mathcal R}^2=2.41\times 10^{-9}$, a
tensor-to-scalar ratio, $r=1$, and a tensor spectral index
$n_T=0$. For the remaining cosmological parameters parameterizing
the background, we use the five-year WMAP
values~\cite{Komatsu:2008hk}.} \label{fig:clbb}
\end{figure}

Using independent Mathematica code, we also checked that the
source function as produced by the modified CMBfast is reliable. We assume
a scale-invariant power spectrum for the tensor perturbations,
$n_{\rm T} = 0$.  The results for a range of masses are shown in
Figure~\ref{fig:clbb}.

In all the plots we drop the quadrupole, because its value depends
on the IR cutoff at low momenta, as follows from the discussion in
section~\ref{zero_mode} (see also section~\ref{disc} for more details). We have used
$\Delta_\mathcal{R}^2=2.41\times 10^{-9}$ and have set the
tensor-to-scalar ratio, $r$, to unity.

In agreement with our estimates, the effect of the mass is rather
mild for masses much below the Hubble rate during recombination
and is present only for very low $\ell$.

For masses approaching the Hubble rate during recombination, the
spectrum is significantly modified up to $\ell\sim 100$, and at
$m_g=3\times 10^4 H_0$ the characteristic plateau at $\ell\lesssim
100$ is fully developed. As we increase the mass further, the
height of the plateau increases up to values of $\mu\equiv
\frac{m_g}{3000 H_0}\approx 25$ but starts to decrease beyond that
in agreement with the oscillations we saw in our semi-analytic
result in Figure~\ref{fig:I2}. The origin of the oscillations is
that depending on the mass  the metric perturbation enters
recombination in different phase.

In agreement with our qualitative arguments summarized in
Figure~\ref{fig:classes}, for masses $\mu\gtrsim 10$ we see that a
transition region appears between the multipole moment $\ell_m$
where the plateau ends and the multipole moment $\ell_0$ where the
massless spectrum is approached.

We are not showing the spectra at higher masses, because the
polarization signal becomes strongly suppressed for $\mu\gg 150$
because of rapid oscillations of the metric during recombination
in agreement with Figure~\ref{fig:I2}.

\begin{figure}[t!]
\begin{center}
\includegraphics[width=6.7in]{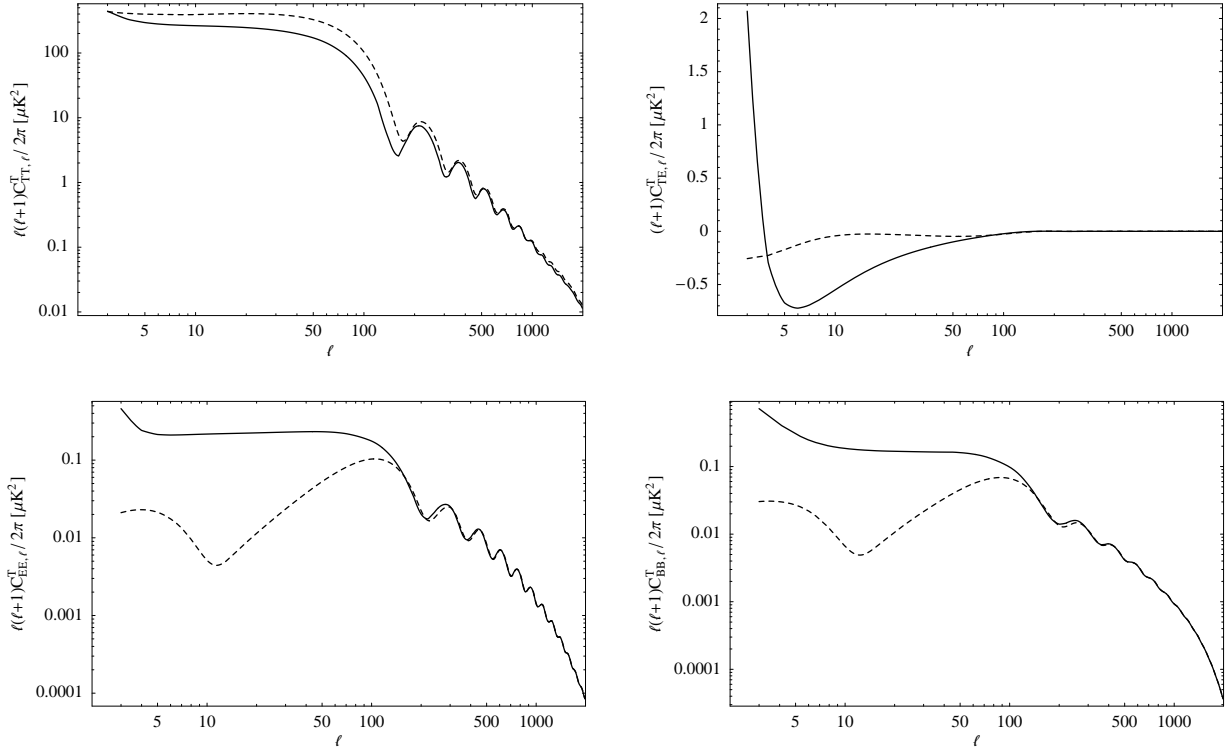}
\end{center}
\caption{This plot shows $T$ (upper left panel), $E$ (lower left), $TE$ (upper right) and
$B$ (lower right) spectra for the massive case with $\mu=10$  (solid line) and
for the massless case (dashed line).
}
\label{fig:4m10}
\end{figure}

So far we have only shown the $C^T_{BB,\ell}$ multipole
coefficients as they are the most interesting from a
phenomenological point of view. In Figure~\ref{fig:4m10}, we show
a comparison of all four CMB spectra (temperature anisotropy, $E$-
and  $B$-type polarization and $TE$ cross-correlations) for
$\mu=10$ and the massless case.

We see that the mass affects  both $E$- and $B$-type polarization
in a rather similar way. The effect of the mass on the temperature
anisotropy is rather mild and the shape of the spectrum for
massive gravity is very similar to the massless case. Unlike
polarization, the temperature anisotropy receives contributions
not only from recombination and reionization, but from all times.
As a result, the contribution at horizon crossing dominates and
one obtains a plateau reflecting the flatness of the primordial
spectrum in both the massive and the massless case (see
\cite{St85} for the analytic expression describing this plateau in
the massless case which, as is seen from the upper left plot in
Figure~\ref{fig:4m10}, produces a good approximation to the
massive case, too). While the polarization spectra look rather
different from the ones in the massless case, one should have in
mind that we only show the tensor contribution to the signal. The
main component for the temperature anisotropy as well as the TE
and EE spectra comes from the scalar perturbations which are
identical to the ones in general relativity. One may still wonder
how large a tensor signal one could tolerate in the massive case
given the existing data and what has already been ruled out. To
this end, we perform a Markov chain Monte Carlo study for a single
value of mass $m_g=3\times 10^4 H_0$, or equivalently $\mu=10$.

We use the publicly available CosmoMC
code~\cite{Lewis:2002ah,cosmomc} to sample the parameter space
together with CAMB~\cite{Lewis:1999bs,camb} to generate the
spectra for a given set of cosmological parameters, and we use a
modified version of the WMAP likelihood code that is now available
on the LAMBDA website~\cite{lambda} to evaluate the likelihood
function for a given spectrum.

In addition to varying the six parameters of the $\Lambda$CDM
model and marginalizing over the Sunyaev-Zeldovich amplitude, we
allow the tensor-to-scalar ratio to vary but keep the mass and the
tensor spectral index fixed. We do not implement the slow-roll
consistency condition but set $n_T=0$ as one may expect the
consistency condition to be modified in these theories, but this
does not significantly change the results.

\begin{figure}[t!]
\begin{center}
\includegraphics[width=6in]{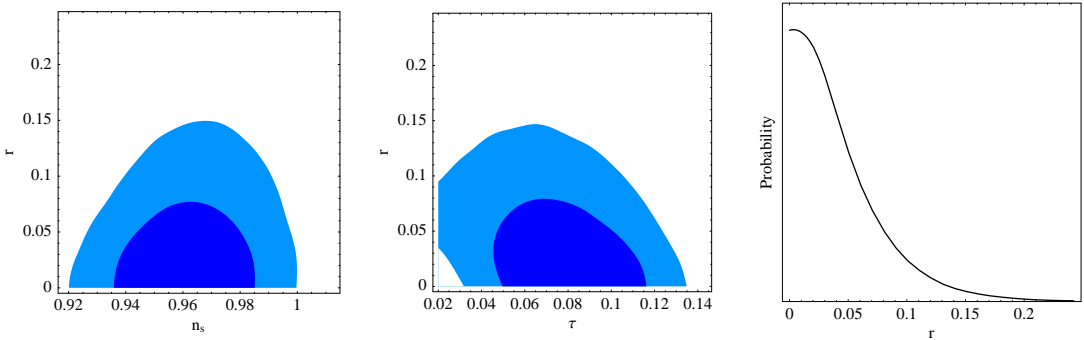}
\end{center}
\caption{These plots show marginalized likelihood plots obtained
from a Markov chain Monte Carlo study of a massive gravity model
with a mass $m_g=3\times 10^4 H_0$,  or equivalently $\mu=10$
using the five-year WMAP data. The dark and light blue contours
correspond to $68\%$ and $95\%$ confidence level, respectively. }
\label{fig:r10}
\end{figure}

We find that the tensor-to-scalar ratio for this value of mass is
constrained to $r<0.11$ at $95\%$ confidence level by the
five-year WMAP data alone. The results are shown in
Figure~\ref{fig:r10}.

Different from the massless case~\cite{Komatsu:2008hk}, where
adding additional data sets like BBN, supernovae, or baryon acoustic oscillations significantly lowers the
allowed tensor-to-scalar ratio, the results do not change much for
the massive case. In the massless case, the reason for the
significant strengthening of the bound is that additional data
sets constraining the baryon or dark matter abundance break a
chain of degeneracies. The main constraint on the tensor-to-scalar
signal in the massless case currently comes from the low-$\ell$ TT
spectrum, where the tensor signal has a plateau as can be seen
{\it e.g} in Figure~\ref{fig:4m10}. Raising $n_S$ lowers the power
in the scalar spectrum at low $\ell$ thus allowing for larger $r$.
On the other hand raising $n_S$ becomes possible only because of a
degeneracy between $\Omega_B$ and $n_S$ so that adding the BBN
priors on the baryon abundance eliminates this possibility and
substantially lowers the bound on the tensor-to-scalar ratio in
the massless case. In the massive case, however, the polarization
data is starting to constrain the model, eliminating the
degeneracy in a different way. Including a BBN prior then does not
significantly lower the bound in this case and all in all the
bound is roughly a factor of two stronger for $m_g=3\times 10^4
H_0$ than it is in the massless case. As one might expect from
looking at the spectra in the massive case, there is now a
degeneracy between the tensor-to-scalar ratio and the optical
depth, however. This is shown in Figure~\ref{fig:r10}. We
interpret these results as telling us that the model is not in
conflict with present data but not much more. In a more serious
analysis, the mass should certainly not be taken as fixed but be
thought of as an unknown parameter that has to be extracted from
the data. We leave a more systematic study for when more data
becomes available.

\section{Discussion}\label{disc}
To summarize, we see that a detection of the CMB B-mode signal
either with Planck or with next generation CMB measurements such
as CMBPol~\cite{Baumann:2008aq}, in addition of opening a new
observational window on inflation,  will also provide a sensitive
probe of the graviton mass. We showed that the most interesting
consequence of the graviton mass for the cosmic microwave
background is probably the characteristic plateau in the $B$-mode
spectrum for multipoles with $\ell\lesssim 100$. This plateau is
most pronounced for masses a few times the Hubble rate at
recombination, but in principle CMB polarization measurements are
capable of constraining the graviton mass down to $m_g^{-1}\sim 10
\mbox{ Mpc}$. Taking into account that large graviton masses
$m_g^{-1}\ll10 \mbox{ kpc}$ lead to a strong suppression of the
tensor contribution to the CMB spectra, we conclude that the
observation of $B$-mode with the conventional inflationary
spectrum would provide by far the tightest bound on the mass of an
elementary particle -- a graviton -- at a level of $m_g\lesssim
10^{-30}$~eV.

Of course, even more exciting would be to find out that a graviton
mass is actually non-zero. It is worth stressing that
gravitational waves after being produced during inflation remain
practically undisturbed throughout the later evolution of the
Universe~\cite{Starobinsky:1979ty} and secondary sources of the
$B$ mode, such as the weak lensing contribution, are negligibly
small at $\ell<100$ \cite{Zaldarriaga:1998ar}. It also appears
extremely hard to mock up the effect of the mass by modifying an
inflationary model or invoking another mechanism generating
gravitational waves such as cosmic strings. Consequently, the
detection of a $B$-mode signal with the shape discussed above
would provide an unambiguous signal of a graviton mass.

We have also seen that in a massive gravity theory superhorizon
tensor perturbations are physical and contribute to the quadrupole
of TT, EE, and TE spectra. As a consequence, the amplitude of
these quadrupoles are IR sensitive and can be significantly
enhanced over the rest of the spectrum. The ratio between the
quadrupole and the rest of the spectrum is model dependent. It
provides a      probe of the total duration of the inflation
provided the tensor perturbations are naturally set to zero at
early times in a given model. This is the case, for instance, if
the graviton mass is not constant during inflation and in the
beginning is bigger than the expansion rate. One way this
possibility may be realized is if the dilatation symmetry
$t\to\lambda t,\; x^i\to\lambda^{-\gamma} x^i$  is not present in
the full theory, but only gets restored during inflation as one
approaches a cosmological attractor discussed in
\cite{Dubovsky:2005dw}. In this particular scenario, the
quadrupoles measure the number of e-folds during the period for
which the graviton is lighter than the Hubble rate.

\section{Acknowledgments}
It is a pleasure to thank Eiichiro Komatsu, Peter Tinyakov and
Matias Zaldarriaga for useful discussions. SD thanks the Theory
Group at the University of Texas for hospitality during the early
stages of this project. RF would like to thank the Stanford
Institute for Theoretical Physics for hospitality during the early
stages of this work and the Aspen Center for Physics during the
late stages. The work of SD and RF has been partially supported by
the National Science Foundation under Grant No. PHY-0455649. AS
was partially supported by the grant RFBR 08-02-00923 and by the
Scientific Programme "Elementary Particles" of the Russian Academy
of Sciences.


\begin{thebibliography}{99}
\bibitem{Fierz:1939ix}
  M.~Fierz and W.~Pauli,
  ``On Relativistic Wave Equations For Particles Of Arbitrary Spin In An
  Electromagnetic Field",
  Proc.\ Roy.\ Soc.\ Lond.\ A {\bf 173}, 211 (1939).

\bibitem{vanDam:1970vg}
H.~van Dam and M.~J.~G.~Veltman, ``Massive And Massless Yang-Mills
And Gravitational Fields'', Nucl.\ Phys.\ B {\bf 22}, 397 (1970).

\bibitem{Zakharov}
V.~I.~Zakharov, "Linearized Gravitation Theory and the Graviton
Mass", JETP Lett. {\bf 12}, 312 (1970)

\bibitem{Vainshtein:1972sx}
  A.~I.~Vainshtein,
  ``To The Problem Of Nonvanishing Gravitation Mass'',
  Phys.\ Lett.\ B {\bf 39}, 393 (1972).

\bibitem{Boulware:1973my}
D.~G.~Boulware and S.~Deser, ``Can Gravitation Have A Finite
Range?'', Phys.\ Rev.\ D {\bf 6}, 3368 (1972).

\bibitem{Arkani-Hamed:2002sp}
N.~Arkani-Hamed, H.~Georgi and M.~D.~Schwartz, ``Effective field
theory for massive gravitons and gravity in theory space'', Annals
Phys.\  {\bf 305}, 96 (2003) [arXiv:hep-th/0210184].

\bibitem{Rubakov:2008nh}
  V.~A.~Rubakov and P.~G.~Tinyakov,
  ``Infrared-modified gravities and massive gravitons'',
  Phys.\ Usp.\  {\bf 51}, 759 (2008)
  [arXiv:0802.4379 [hep-th]].

\bibitem{Arkani-Hamed:2003uy}
N.~Arkani-Hamed, H.~C.~Cheng, M.~A.~Luty and S.~Mukohyama, ``Ghost
condensation and a consistent infrared modification of gravity'',
JHEP {\bf 0405}, 074 (2004) [arXiv:hep-th/0312099].

\bibitem{Rubakov:2004eb}
V.~Rubakov, ``Lorentz-violating graviton masses: Getting around
ghosts, low strong coupling scale and VDVZ discontinuity'',
arXiv:hep-th/0407104.

\bibitem{Dubovsky:2004sg}
S.~L.~Dubovsky, ``Phases of massive gravity'', JHEP {\bf 0410},
076 (2004) [arXiv:hep-th/0409124].

\bibitem{Dubovsky:2004ud}
  S.~L.~Dubovsky, P.~G.~Tinyakov and I.~I.~Tkachev,
  ``Massive graviton as a testable cold dark matter candidate'',
  Phys.\ Rev.\ Lett.\  {\bf 94}, 181102 (2005)
  [arXiv:hep-th/0411158].

\bibitem{Dubovsky:2005dw}
  S.~L.~Dubovsky, P.~G.~Tinyakov and I.~I.~Tkachev,
  ``Cosmological attractors in massive gravity'',
  Phys.\ Rev.\  D {\bf 72}, 084011 (2005)
  [arXiv:hep-th/0504067].


\bibitem{Bebronne:2007qh}
  M.~V.~Bebronne and P.~G.~Tinyakov,
  ``Massive gravity and structure formation'',
  Phys.\ Rev.\  D {\bf 76}, 084011 (2007)
  [arXiv:0705.1301 [astro-ph]].

\bibitem{Dubovsky:2007zi}
  S.~Dubovsky, P.~Tinyakov and M.~Zaldarriaga,
  ``Bumpy black holes from spontaneous Lorentz violation'',
  JHEP {\bf 0711}, 083 (2007)
  [arXiv:0706.0288 [hep-th]].

\bibitem{Berezhiani:2007zf}
  Z.~Berezhiani, D.~Comelli, F.~Nesti and L.~Pilo,
  ``Spontaneous Lorentz breaking and massive gravity'',
  Phys.\ Rev.\ Lett.\  {\bf 99}, 131101 (2007)
  [arXiv:hep-th/0703264].

\bibitem{Berezhiani:2008nr}
  Z.~Berezhiani, D.~Comelli, F.~Nesti and L.~Pilo,
  ``Exact Spherically Symmetric Solutions in Massive Gravity'',
  JHEP {\bf 0807}, 130 (2008)
  [arXiv:0803.1687 [hep-th]].

\bibitem{pulsars}
  J.~H.~Taylor,
  ``Binary pulsars and relativistic gravity'',
  Rev.\ Mod.\ Phys.\  {\bf 66}, 711 (1994).

\bibitem{Pshirkov:2008nr}
  M.~Pshirkov, A.~Tuntsov and K.~A.~Postnov,
  ``Constraints on the massive graviton dark matter from pulsar timing and
  precision astrometry'',
  Phys.\ Rev.\ Lett.\  {\bf 101}, 261101 (2008)
  [arXiv:0805.1519 [astro-ph]].

\bibitem{Baumann:2008aq}
  D.~Baumann {\it et al.}  [CMBPol Study Team Collaboration],
  ``CMBPol Mission Concept Study: Probing Inflation with CMB Polarization,''
  AIP Conf.\ Proc.\  {\bf 1141}, 10 (2009)
  [arXiv:0811.3919 [astro-ph]].

\bibitem{seljak}
 M.~Zaldarriaga and U.~Seljak,
  ``An All-Sky Analysis of Polarization in the Microwave
  Background'',
  Phys.\ Rev.\ D {\bf 55}, 1830 (1997)
  [arXiv:astro-ph/9609170].




\bibitem{polnarev}
 A.~G.~Polnarev,
 "Polarization and anisotropy induced in the microwave background
 by cosmological gravitational waves",
 Sov.\ Astron.\ {\bf 29}, 607 (1985).


\bibitem{Crittenden:1993ni}
  R.~Crittenden, J.~R.~Bond, R.~L.~Davis, G.~Efstathiou and P.~J.~Steinhardt,
  ``The Imprint of gravitational waves on the cosmic microwave
  background'',
  Phys.\ Rev.\ Lett.\  {\bf 71}, 324 (1993)
  [arXiv:astro-ph/9303014].


\bibitem{Weinberg:2006hh}
  S.~Weinberg,
  ``A no-truncation approach to cosmic microwave background
  anisotropies'',
  Phys.\ Rev.\  D {\bf 74}, 063517 (2006)
  [arXiv:astro-ph/0607076].

\bibitem{Baskaran:2006qs}
  D.~Baskaran, L.~P.~Grishchuk and A.~G.~Polnarev,
  ``Imprints of relic gravitational waves in cosmic microwave background
  radiation'',
  Phys.\ Rev.\  D {\bf 74}, 083008 (2006)
  [arXiv:gr-qc/0605100].

\bibitem{AZakharov}
A.~V.~Zakharov, "Effect of collisionless particles of the growth
of gravitational perturbations in an isotropic world", Sov.\
Phys.\ - JETP {\bf 50}, 221 (1979).

\bibitem{Weinberg:2003ur}
  S.~Weinberg,
  ``Damping of tensor modes in cosmology'',
  Phys.\ Rev.\  D {\bf 69}, 023503 (2004)
  [arXiv:astro-ph/0306304].

\bibitem{Komatsu:2008hk}
  E.~Komatsu {\it et al.}  [WMAP Collaboration],
  ``Five-Year Wilkinson Microwave Anisotropy Probe (WMAP)
  Observations:Cosmological Interpretation'',
  Astrophys.\ J.\ Suppl.\  {\bf 180}, 330 (2009)
  [arXiv:0803.0547 [astro-ph]].

\bibitem{St85}
 A.~A.~Starobinsky, "Cosmic background anisotropy induced by
 isotropic, flat-spectrum gravitational-wave perturbations",
 Sov.\ Astron.\ Lett.\ {\bf 11}, 133 (1985).

\bibitem{Zaldarriaga:1995gi}
  M.~Zaldarriaga and D.~D.~Harari,
  ``Analytic approach to the polarization of the cosmic microwave background in
  flat and open universes'',
  Phys.\ Rev.\  D {\bf 52}, 3276 (1995)
  [arXiv:astro-ph/9504085].


\bibitem{Pritchard:2004qp}
  J.~R.~Pritchard and M.~Kamionkowski,
  ``Cosmic microwave background fluctuations from gravitational waves: An
  analytic approach'',
  Annals Phys.\  {\bf 318}, 2 (2005)
  [arXiv:astro-ph/0412581].

\bibitem{Keating:2006zy}
  B.~G.~Keating, A.~G.~Polnarev, N.~J.~Miller and D.~Baskaran,
  ``The Polarization of the Cosmic Microwave Background Due to Primordial
  Gravitational Waves'',
  Int.\ J.\ Mod.\ Phys.\  A {\bf 21}, 2459 (2006)
  [arXiv:astro-ph/0607208].


\bibitem{Flauger:2007es}
  R.~Flauger and S.~Weinberg,
  ``Tensor Microwave Background Fluctuations for Large Multipole
  Order'',
  Phys.\ Rev.\  D {\bf 75}, 123505 (2007)
  [arXiv:astro-ph/0703179].

\bibitem{bateman}
  Bateman Manuscript Project,
  ``Higher Transcendental Functions, Volume 2'',
  {\it New York, McGraw Hill, 1953-55}

\bibitem{PS96}
 D.~Polarski and A.~A.~Starobinsky,
 "Semiclassicality and Decoherence of Cosmological Perturbations",
 Class.\ Quant.\ Grav. {\bf 13}, 377 (1996)
 [arXiv:gr-qc/9504030].

\bibitem{baskopolnarev}
 M.~M.~Basko and A.~G.~Polnarev,
 "Polarization and anisotropy of the primordial radiation in an
 anisotropic universe",
 Sov.\ Astron.\ {\bf 24}, 268 (1980).

\bibitem{cmbfast} The code is avaliable at http://www.cmbfast.org/

\bibitem{Seljak:1996is}
  U.~Seljak and M.~Zaldarriaga,
  ``A Line of Sight Approach to Cosmic Microwave Background
  Anisotropies'',
  Astrophys.\ J.\  {\bf 469}, 437 (1996)
  [arXiv:astro-ph/9603033].

\bibitem{Lewis:2002ah}
  A.~Lewis and S.~Bridle,
  ``Cosmological parameters from CMB and other data: a Monte-Carlo
  approach'',
  Phys.\ Rev.\  D {\bf 66}, 103511 (2002)
  [arXiv:astro-ph/0205436].

\bibitem{cosmomc} The code is avaliable at http://cosmologist.info/cosmomc/

\bibitem{Lewis:1999bs}
  A.~Lewis, A.~Challinor and A.~Lasenby,
  ``Efficient Computation of CMB anisotropies in closed FRW
  models'',
  Astrophys.\ J.\  {\bf 538}, 473 (2000)
  [arXiv:astro-ph/9911177].

\bibitem{camb} The code is avaliable at http://camb.info/

\bibitem{lambda} http://lambda.gsfc.nasa.gov/

\bibitem{Starobinsky:1979ty}
 A.~A.~Starobinsky, "Spectrum of relict gravitational radiation
 and the early state of the universe", JETP Lett.\ {\bf 30}, 682 (1979).

\bibitem{Zaldarriaga:1998ar}
  M.~Zaldarriaga and U.~Seljak,
  ``Gravitational Lensing Effect on Cosmic Microwave Background
  Polarization'',
  Phys.\ Rev.\  D {\bf 58}, 023003 (1998)
  [arXiv:astro-ph/9803150].

\end{thebibliography}
\end{document}